# Symmetry Breaking of Frequency Comb in Varying Normal Dispersion Fiber Ring Cavity

Muhammad Imran Afzal, Kamal Alameh, Senior Member, IEEE and Yong Tak Lee

*Abstract*—We build on a previously reported frequency comb of mode spacing 0.136 nm in a fiber ring cavity of varying normal dispersion [1], to generate, for the first time, a frequency comb of mode spacing 0.144 nm centered at 978.544 nm to demonstrate the symmetry-breaking. By controlling the birefringence of the optical cavity through fiber stretching and polarization control, the spacing of the comb lines increases from 0.136 nm to 0.144 nm, and this small change in mode spacing generates very different spectral symmetry-breaking in the frequency comb relative to the frequency comb of mode spacing 0.136 nm. Interestingly, non-uniform depletion of primary modes is also observed. The experimental results are an important contribution in the continuing effort of understanding the dynamics of frequency combs involving large number of modes, nontrivial nonlinear waves and deterministic chaos.

## I. Introduction

NONLINEAR modes generation and phase-locking in normal dispersion regimes exhibit modulation instability or Cross-Phase-Modulation (XPM) [2], which results in spontaneous noise free operation, thus offering unique advantages over anomalous dispersion, such as mode purity and stability. Therefore, true richness of XPM in dissipative optical fiber cavities is observable in normal dispersion regimes, where phase-locking and amplification could be achieved through Dissipative-four-Wave-Mixing (DFWM) [3]. Typically, to tune and phase-lock the cavity wavelength in a fiber cavity, an external device, such as an interferometer, a tunable filter, or a Fiber Bragg Grating (FBG) is used [4−8]. However, such an external device degrades the performance of the cavity because it (i) increases the cavity loss, (ii) introduces additional noise and (iii) limits the wavelength tuning. On the other hand, controlling the birefringence of the fiber cavity is crucial to achieve passive phase locking for any mode spacing [9-10]. Tuning through birefringence control also produces spectral symmetry-breaking [9-13]. However, symmetry-breaking is also achieved through delay coupling of lasers [14]. In fact, effective frequency spacing variation and delay-coupling of laser modes are crucial initial conditions for the realization of deterministic chaotic-like optical structures [10-18]. Particularly, symmetry-breaking is under intensive investigation involving generation and controlling of non-trivial nonlinear waves such as Rogue waves [19] and artificial event horizons [20-21].

Recently, the authors have reported a frequency comb architecture based on the use of self-induced modulational-instability in a passive fiber ring cavity of varying normal dispersion. By simply controlling the birefringence of its fiber ring cavity, optical frequency comb of spacing 0.136 nm has been successfully generated [1]. In this paper, we build on this reported architecture, however, we specifically investigate a new mode spacing tuning mechanism based on varying the birefringence of the cavity through mechanical fiber stretching in conjunction with polarization control inside the cavity. Mechanical fiber stretching is crucial for inducing large birefringence variation, while polarization control is used for achieving small birefringence variation that enables the generation of frequency combs of variable line spacings. We particularly observe frequency combs of mode spacings 0.136 and 0.144 nm exhibiting symmetry-breaking in the modulation instability spectrum. In particular, the novel contribution of the frequency comb architecture reported in this paper is attributed to its new complex optical cavity, which is capable of generating a relatively large number of modes with spectral symmetry-breaking, achieved through birefringence control. This macro-level symmetry-breaking characteristics and observation of non-uniformly depletion of primary modes, make our experimental results particularly distinguished from previously reported architectures with symmetry-breaking features [10-18]. Some of the recent studies [21-22] identify the significance of depletion of nonlinear primary modes to produce the symmetry breaking.

## II. Experimental Setup

The experimental setup used to demonstrate the concept of the symmetry-breaking frequency comb architecture is shown in Fig. 1. The fiber ring cavity had a total length of 604 cm, which was made up by coupling segments of passive fibers, namely, Hi1060, Hi980, SMF-28e, and SM600 of different core diameters and dispersion values at 980 nm of ~ −53, ~ −63, ~ −63, and ~ −75 ps/nm/km, respectively. The calculated Group Velocity Dispersion (GVD) of the cavity was 0.18 ps2 at 980 nm.

This work was partially supported by the "Systems biology infrastructure establishment grant" and "Asian Laser Center program" provided by GIST. (Corresponding author: Y. T. Lee.)

M. I. Afzal is with Advanced Photonics Research Institute, Gwangju Institute of Science and Technology, Gwangju 500-712, Korea.

K. Alameh is with the Centre of Excellence for MicroPhotonic Systems, Electron Science Research Institute, Edith Cowan University, Joondalup, WA 6027, Australia (e-mail: k.alameh@ecu.edu.au).

Y. T. Lee is with the School of Information and Communications, Department of Physics and Photon Science, and Advanced Photonics Research Institute, Gwangju Institute of Science and Technology, Gwangju 500-712, Korea (e-mail: ytlee@gist.ac.kr).

For a pump power of 130 mW, the optical power inside the cavity after one round trip was estimated to be 83.94 mW and the coupling loss power was 46.06 mW. The ring cavity was pumped with a multimode pump laser of center wavelength ~978 nm. To realize a ring cavity with varying dispersion, four segment fibers of different cores and normal dispersion values were used. The optical signals in the cavity along with dynamically controlling the birefringence resulted in spectral broadening and produced self-induced instability at low power, as reported in [1].

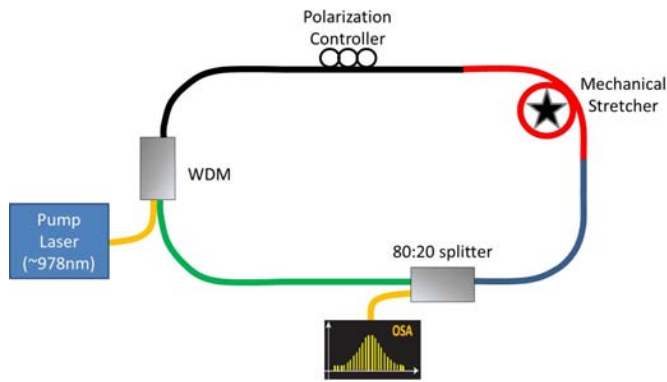

Fig. 1. Experimental setup to demonstrate the concept of the symmetry-breaking frequency comb architecture: The 604 cm long fiber loop consisted of four segments of fibers: Hi 1060 (black 273 cm), SM600 (red 108 cm), HI 980 (blue 104 cm) and SMF-28e (green 119 cm). PC: Polarization Controller.

## III. RESULTS AND DISCUSSIONS

Around a pump power of 130 mW, unstable modes appeared over a broadened spectrum around a peak of 978.444 nm. By carefully controlling the birefringence of the optical cavity, the unstable mode 978.580 nm was stabilized along with 978.444 nm and phase matched, as a result of the Cross-Phase-Modulation (XPM) of optical signals propagating in the cavity. With the fundamental modes being phase matched with higher order modes weakly modulated stokes and antistokes modes were amplified through dissipative four-wave mixing, leading to the formation of a frequency comb of spacing of 0.136 nm. By further increasing the pump power up to 215 mw, the modes were amplified forming clearly visible frequency comb particularly the Stokes modes. This process the power levels of the nonlinear primary modes 978.444 and 978.580 nm from 0 to −11.92 dBm and −5 to −10.09 dBm, respectively, as shown in Figs 2(a) and 2(b). Amplification of comb modes at

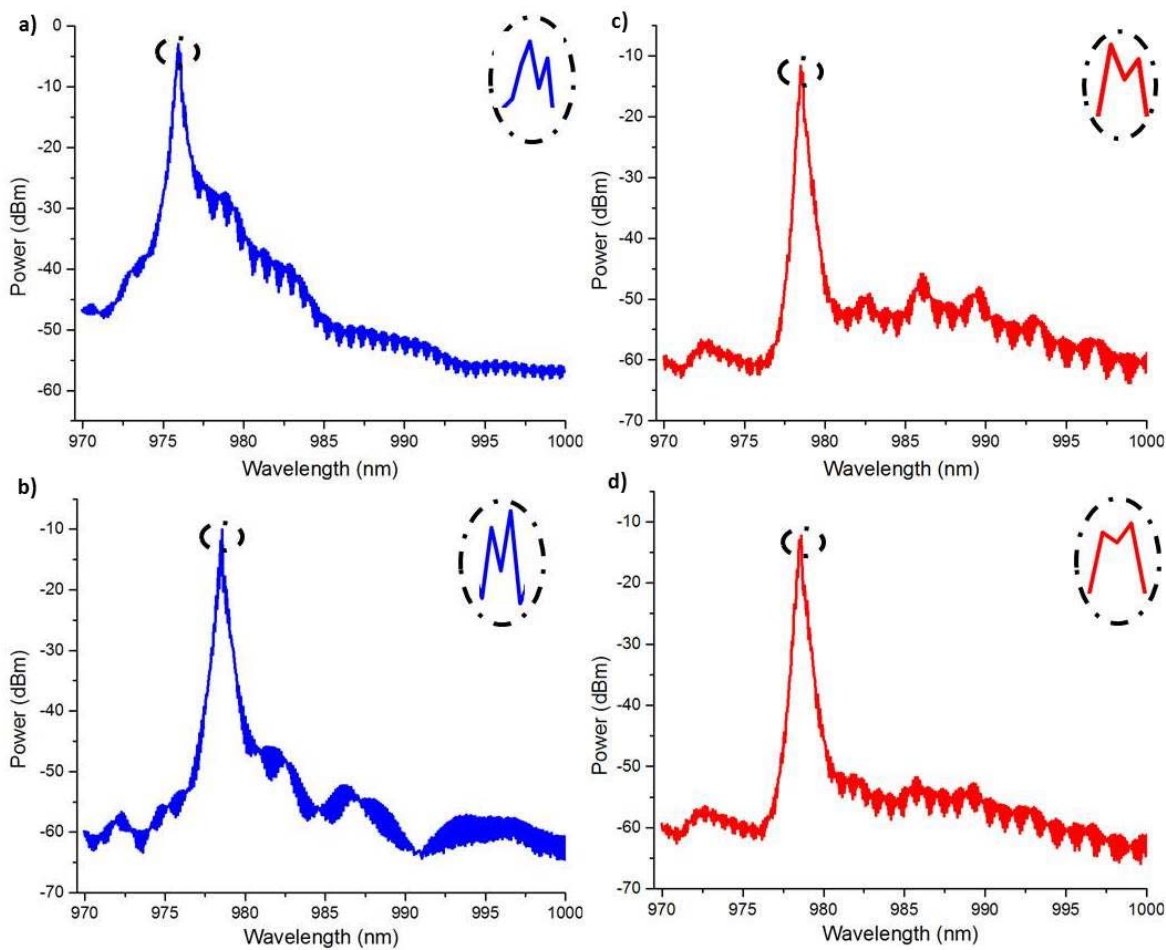

Fig. 2. Generated frequency combs (a) spacing = 0.136 nm and pump power ~130 mW, (b) spacing = 0.136 nm and pump power ~215 mW, (c) spacing = 0.144 nm and pump power ~130 mW, and (d) spacing = 0.144 nm and pump power ~180 mW. Insets: Zoomed primary modes exhibit depletion of power.

of the

primary modes were the evidence of passive phase-locking through (DFWM). High asymmetry was observed over the spectral range 985 to 990 nm while relatively low asymmetry was observed over the range 990 to 1000 nm.

To achieve a mode spacing 0.144 nm, we used the same experimental setup. We restarted the experiment with a pump power of 130 mw and generated unstable modes. By carefully controlling the birefringence of the cavity using a mechanical stretcher and a polarization controller, unstable primary modes at 978.472 and 978.616 nm were stabilized and phase matched due to XPM. As soon as the phase-matching took place, the Stokes and anti-Stokes modes started to generate and nonlinearly amplify through DFWM as shown in Fig. 2(c).

By increasing the pump power from ~130 mW to ~180 mW, the Stokes and anti-Stokes modes were amplified, however, with a less gain, in comparison with that experienced by the 0.136 nm-spacing frequency comb. Interestingly, a smooth spectrum from 985 nm to 990 nm was clearly observed, as shown in Fig. 3(d).

Consequently, the power of the nonlinear primary mode at 978.472 nm was depleted from −11.66 dBm to −13.06 dBm while the mode at 978.616 nm was depleted from −12.88 dBm to −12.24 dBm, as shown in Fig. 2(c) and 2(d). Particularly, the zoomed-in spectra in the insets of Figs 2(a-d) illustrate the power depletion and symmetry-breaking of the primary modes. By further increasing the pump power, no amplification or spectrum smoothening were observed. This is attributed to the complex dynamics of the cavity, which do not allow further power transfer to the Stokes and anti-Stokes modes or smoothness of the spectrum. This is also evident from the fact that no further depletion of the power levels of the nonlinearly generated primary modes was observed. Power depletion of nonlinear primary modes has particular importance in generation and amplification of frequency combs and generation of artificial event horizons [21-22]. Better smoothening of the spectrum was observed over the spectral range 985 nm to 990 nm as evident from Figs 3(c) and 3(d). On the other hand, for the 0.136 nm-spacing frequency comb, large amplification as well as asymmetry was observed, while for the lower frequency modes (990 to 1000 nm), relatively a smooth spectrum was observed as evident from

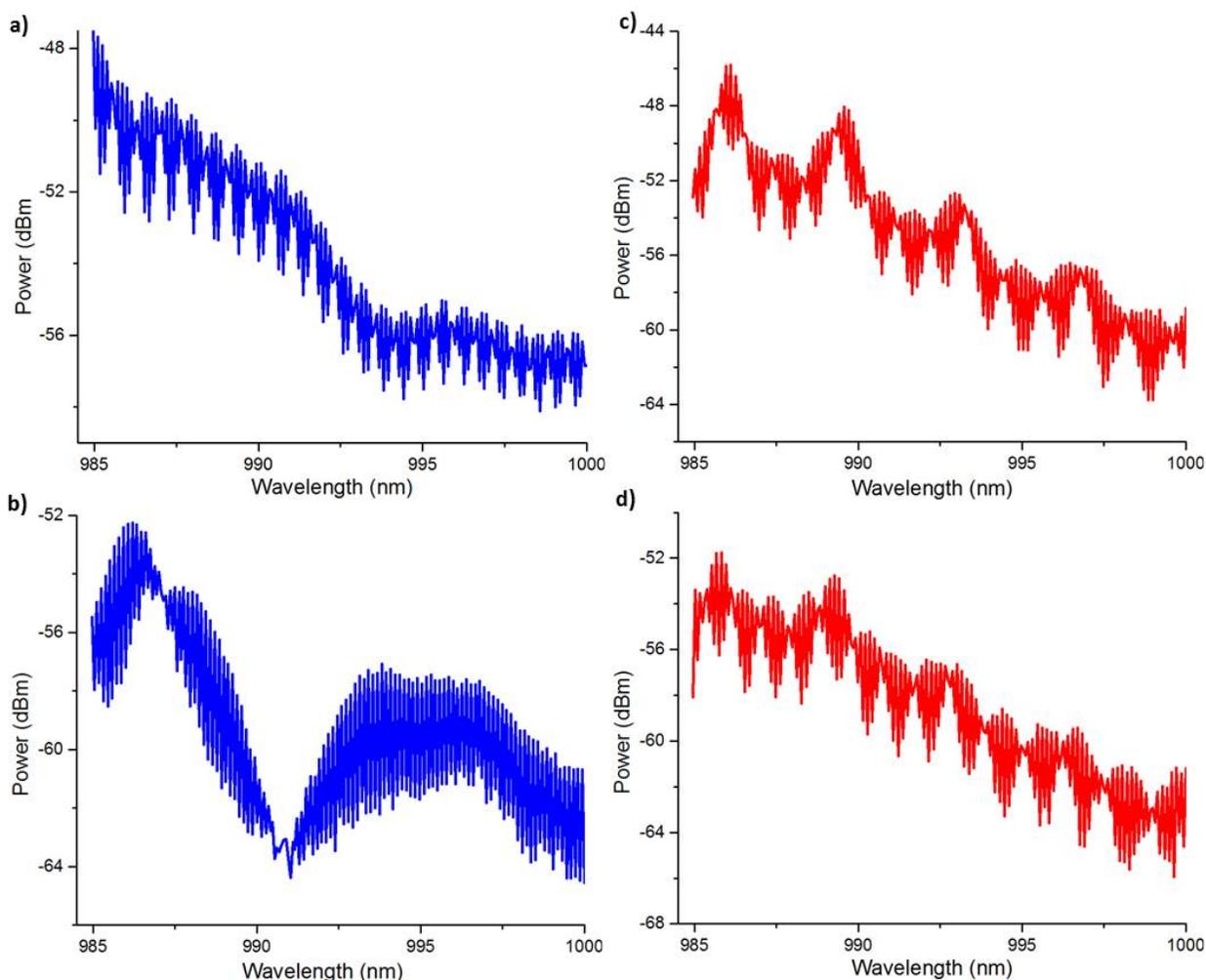

Fig. 3. Generated frequency combs with and without symmetry-breaking and amplification over the spectral range (985 to 1000 nm) (a) line spacing = 0.136nm and pump power ~130 mW, (b) line spacing = 0.136nm and pump power increased to ~215 mW, (c) line spacing = 0.144 nm and pump power ~130 mW, (d) line spacing = 0.144nm and pump power increased to ~180 mW, displaying very small amplification but substantial increase in symmetry, particularly from 985 to 990 nm.

Figs 3(a) and 3(b). The observed symmetry-breaking was primarily due to phase matching between the primary modes and higher order modes in the optical cavity, which comprises optical fibers of different properties.

## IV. CONCLUSIONS

The present study has demonstrated the generation of frequency combs with symmetry-breaking through complex optical cavity comprising (i) four optical fibers of different properties connected in series, (ii) a multimode pump laser, and (iii) birefringence control mechanism for phase matching control. Mode spacings of 0.136 nm and 0.144 nm have been experimentally attained, together with spectral symmetry-breaking, which is attributed to asymmetric power transfer to comb lines through DFWM processes in the complex optical cavity. This explanation is also supported by the observation of non-uniform depletion of nonlinear primary modes. The experimental results open the way for further research and development of more complex frequency combs of arbitrary spectral profiles, where amplification of individual or small group of modes can be controlled via simple optical control mechanisms. Our present study is of particularly important to understand the universal dynamics of optical frequency combs [22] and nontrivial nonlinear waves [19−21] in terms of sensitive dependence on change in initial frequency spacing or time-delay [22], particularly in the presence of large number of phase-locked modes in the varying dispersion environment and could provide clues for further theoretical and experimental investigations of the dynamics of very large chaotic systems [23].